\documentclass[prl,amsmath]{revtex4-1}
\usepackage[pdftex]{graphicx} 
\usepackage{color}
\usepackage{bm}
\usepackage{epsfig}
\usepackage{latexsym}
\usepackage{nameref,hyperref}
\begin{document}
\newcommand{\be}{\begin{equation}}
\newcommand{\ee}{\end{equation}}
\newcommand{\bea}{\begin{eqnarray}}
\newcommand{\eea}{\end{eqnarray}}
\newcommand{\RN}[1]{\textup{\uppercase\expandafter{\romannumeral#1}}}

\title{Effect of receptor clustering on chemotactic performance of Escherichia coli: sensing versus adaptation}
\author{Shobhan Dev Mandal and Sakuntala Chatterjee}
\affiliation{Department of Theoretical Sciences, S. N. Bose National Centre for Basic Sciences, Block JD, Sector 3, Salt Lake, Kolkata 700106, India.}
\begin{abstract}
We show how the competition between sensing and adaptation can result in a performance peak in E.coli chemotaxis using extensive numerical simulations in a detailed theoretical model. Receptor clustering amplifies the input signal coming from ligand binding which enhances chemotactic efficiency. But large clusters also induce large fluctuations in total activity since the number of clusters go down. The activity and hence the run-tumble motility now gets controlled by methylation levels which are part of adaptation module, rather than ligand binding. This reduces chemotactic efficiency.
\end{abstract}
\maketitle

With the advent of sophisticated techniques to measure single-cell response in experiments \cite{sourjik2002receptor, turner2000real}, an important question has emerged: how behavior of a cell is affected by the fluctuations present in the intracellular biochemical reaction network \cite{elowitz2002stochastic, raj2008nature, rao2002control, lan2016information}. In this paper we address this question for Escherichia coli chemotaxis, one of the best characterized systems in biology \cite{berg2008coli}.

The chemotaxis describes the migration tendency of the Escherichia coli cell towards the region of higher nutrient concentration. The underlying biochemical network has two main modules, sensing and adaptation, which are coupled to each other through the activity of the transmembrane chemoreceptors. The receptor activity changes with binding of the receptor to the nutrient ligand molecules, and with methylation. There are a few thousand receptors in a cell and they show strong cooperativity where the neighboring receptors form clusters or `teams' and switch between active and inactive states in unison. This helps in amplification of the input signal coming from ligand binding and allows the cell to respond to even weak gradient of nutrient concentration \cite{frank2016networked, duke1999heightened, bray1998receptor}.

 In recent experiments involving single cell FRET measurements it was observed that receptor clustering results in surprisingly large activity fluctuations inside a cell \cite{colin2017multiple, keegstra2017phenotypic} even in the absence of methylation noise. This observation was striking since methylation was long believed to be the most important source of noise in a chemotaxis network \cite{korobkova2004molecular, matthaus2011origin, flores2012signaling, dev2018optimal, tu2005white}. The experiments in Refs. \cite{colin2017multiple, keegstra2017phenotypic} showed that receptor clustering is an independent and equally important noise source in the pathway. The immediate and important question here is, how this newly found noise source is related to the chemotactic performance of the cell.

In this work, we address this question within a detailed theoretical model and find that there is an optimum size of the receptor cluster at which the chemotactic performance is at its best. Since receptor clustering amplifies the input signal coming from ligand binding, it is expected to enhance the cell performance \cite{frank2016networked, duke1999heightened, bray1998receptor}. However, when clusters become significantly large, the total number of clusters goes down proportionately. The total activity of the cell, which is the sum of activity of all the clusters, starts showing large fluctuations since the sum is now performed over a small number of signaling teams (also see Sec. $3$ of Ref. \cite{sup_mat}). When the activity gets too high or too low, the adaptation comes into play and the receptor methylation level undergoes large change to restore the activity to its mean value. Our data show that the total activity which controls the run-and-tumble motility of the cell is guided by methylation, rather than ligand binding for large receptor clusters. This reduces the chemotactic efficiency of the cell and its performance goes down.

Our study brings out a fundamentally important point: how competition between sensing and adaptation may result in a performance peak. We demonstrate this by monitoring several different quantities as measures of performance. In the presence of a spatially varying nutrient concentration profile we define a good chemotactic performance by measuring how fast the cell is able to climb up the gradient, or how strongly it is able to localize itself in the nutrient-rich regions \cite{dev2015optimal, dev2018optimal}. A good performance implies a strong ability of the cell to distinguish between regions with high and low nutrient concentration. We find that for an optimal size of the receptor cluster this ability is most pronounced. Interestingly, our conclusion remains valid even  when the cell is tethered and is not moving around using run-tumble motility. In this case we define the performance by the differential response of the cell when the nutrient level at its location is increased or decreased. The rotational bias of the flagellar motors shows maximum difference between the ramped up and ramped down inputs at a specific size of the receptor cluster.

{\it Model:} In an E. coli cell the chemoreceptors pair up to form homodimers and three such homodimers form a trimer of dimers (TD) \cite{liu2012molecular, briegel2012bacterial}. In our description, a signaling team of size $n$ contains $n$ number of TDs. The free energy difference (in units of $K_B T$) between the active and inactive states of a dimer is calculated according to Monod-Wyman-Changeux  model \cite{mello2005allosteric, keymer2006chemosensing, monod1965nature}:
\be
\epsilon [m,c(x)]= 1+ \log\frac{1+c(x)/K_{min}}{1+c(x)/K_{max}} -m   \label{eq:ep}
\ee
where $c(x)$ is the nutrient concentration at the cell location $x$ and $m$ is the methylation level of the dimer which can take integer values between $0$ and $8$. The constants $K_{min}$ and $K_{max}$ set the range within which a chemical concentration can be sensed by the cell. The total free energy of the cluster is the sum of free energy of the individual dimers. All dimers in a cluster change their activity states simultaneously and the transition probability depends on the cluster free energy \cite{sup_mat}.

 The methylation level of a dimer is controlled by methylating enzyme CheR and demethylating enzyme CheB-P. A dimer can bind to one enzyme molecule at a time. An inactive dimer gets methylated by CheR and the probability to find it in active state increases. On the other hand, an active dimer gets demethylated by CheB-P and its activity decreases. Unphosphorylated CheB receives its phosphate group from autophosphorylation of active receptors. This constitutes a negative feedback in the reaction network and is responsible for adaptation.  Autophosphorylation of active receptors also supplies phosphate group to another protein CheY and the resulting CheY-P binds to the flagellar motors and induces tumbling in the cell motion. A high value of total activity implies large tumbling probability.

However, the number of enzyme molecules is far too low compared to the number of dimers in a cell \cite{li2004cellular} and it takes a long time for a dimer to bind to an unbound enzyme molecule in cell cytoplasm \cite{schulmeister2008protein}. To reconcile the low abundance of enzyme molecules with near-perfect adaptation of the cell \cite{berg1975transient, goy1977sensory}, few mechanisms like `brachiation' or `assistance neighborhood' have been proposed \cite{levin2002binding, endres2006precise, hansen2008chemotaxis} and experimentally verified \cite{kim2002dynamic, li2005adaptational} which allow a single bound enzyme to modify the methylation level of more than one dimers before it unbinds and returns to the cell cytoplasm \cite{keymer2006chemosensing, le1997methylation, hansen2008chemotaxis}. We include a flavor of this mechanism in our model. A complete description of our model and other simulation details can be found in Ref \cite{sup_mat}. We perform Monte Carlo simulations on this model in one and two spatial dimensions.  We present the data for two dimensions below and include those for one dimension in \cite{sup_mat}.

\begin{figure}
\includegraphics[scale=1.2]{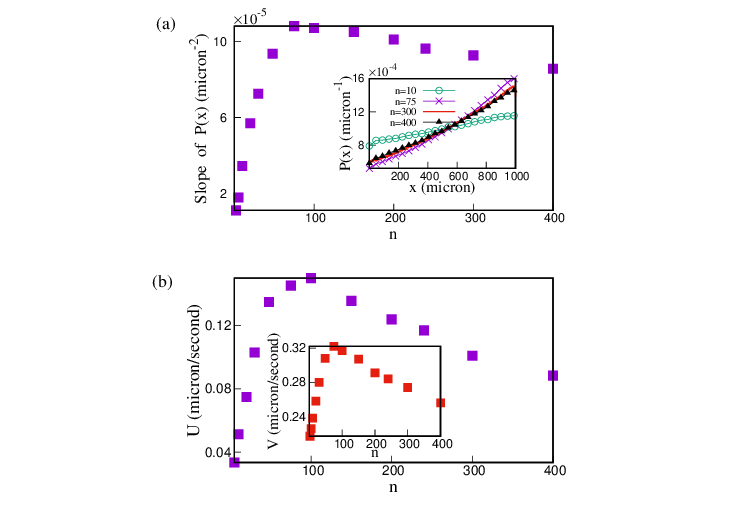}
\caption{Peak in localization and drift velocity as a function of receptor cluster size. (a) The $x$-position distribution of the cell shows steepest variation at an optimum $n$. Inset shows form of $P(x)$ for few representative $n$ values. (b) Chemotactic drift velocity measured from net displacement in a run (inset) and net displacement in a fixed time-interval $T=40$ $s$  (main plot), both show peak for a specific $n$. We have used a linearly varying nutrient concentration profile here. Each data point has been averaged over at least $10^7$ histories. The simulation parameters are given in Table S1 in \cite{sup_mat}.}
\label{fig:pxv}
\end{figure}

{\it Performance peak at an optimal size of receptor cluster:} For a swimming cell with a linearly varying $c(x)$, the steady state position distribution $P(x)$ of the cell also assumes an almost linear form (see inset of Fig. \ref{fig:pxv}a). Clearly, a good performance implies a steep slope of $P(x)$. In Fig. \ref{fig:pxv}a (main plot) we plot this slope as a function of receptor cluster size and find a peak. A related quantity $\langle C \rangle = \int P(x) c(x) dx$ which gives the average nutrient amount experienced by the cell, is often used to characterize performance when $c(x)$ or $P(x)$ is not linear \cite{dev2018optimal, flores2012signaling, celani2010bacterial}. We find a similar peak in $\langle C \rangle$ also (data not shown here). Chemotactic drift velocity $V$ measures how fast the cell climbs up the concentration gradient and a large $V$ implies a good performance. To extract $V$ from the run-and-tumble trajectory of the cell we measure the mean value of net displacement of the cell in a run and divide it by the mean run duration \cite{de2004chemotaxis, chatterjee2011chemotaxis, pankratova2018chemotactic, samanta2017predicting, dev2018optimal}. We present our data in Fig. \ref{fig:pxv}b  inset which shows a pronounced peak. Another possible way to measure the drift velocity is from the net displacement in a fixed time interval $T$ and divide that by $T$. In the  main plot of Fig. \ref{fig:pxv}b we show the plot for this quantity, denoted as $U$ and find a similar peak. 

\begin{figure}
\includegraphics[scale=1.2]{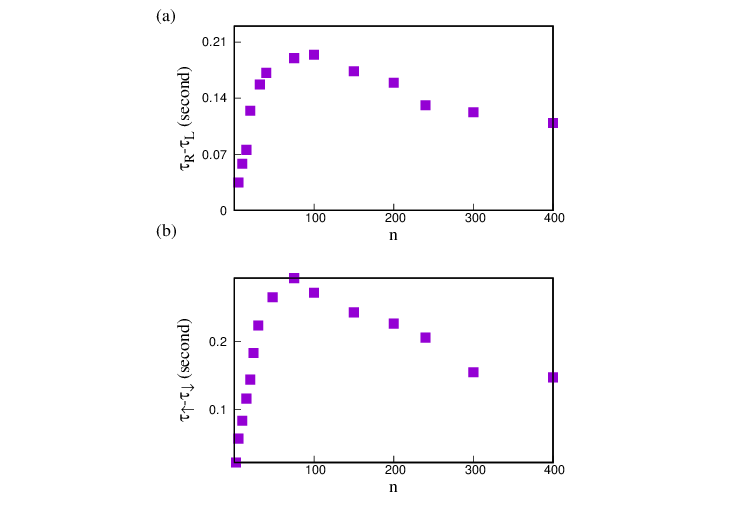}
\caption{Motor response of the cell shows highest sensitivity at a specific size of receptor cluster. (a) For a swimming cell, the mean first passage time to the tumble mode for uphill run $(\tau_R$) and downhill run $(\tau_L$) shows largest difference at a particular $n$. (b) For a tethered cell in CCW mode, the mean first passage time to CW mode when the nutrient level is ramped up ($\tau_{\uparrow}$) and  ramped down ($\tau_\downarrow$) at a rate $0.1$ $\mu M/s$ shows the largest difference at a specific $n$. All data have been averaged over at least $10^6$ histories. The simulation parameters are as in Fig. \ref{fig:pxv}.}
\label{fig:ftum}
\end{figure}

At the core of chemotactic sensing lies the differential behavior of the cell when the nutrient level in its environment goes up or down. This difference should be large for a good performance. When a cell is running in the direction of increasing nutrient concentration, its tumbling rate decreases and the run is extended. Similarly, for a run towards lower nutrient level, the tumbling rate increases and the run is shortened. We measure the time till the first tumble during an uphill run and a downhill run and plot their difference in  Fig. \ref{fig:ftum}a. This difference shows a peak at a specific size of the receptor cluster. Interestingly, we can use a similar measure to quantify performance for a tethered cell as well, which is more commonly used in experiments. In this case we apply a nutrient concentration that is increasing (decreasing) linearly with time while the flagellar motors are rotating in the counter clockwise (CCW) direction. We measure the average time till the transition to clockwise (CW) rotation mode. In order to compare with the swimming cell, we change the nutrient level at the same rate as that experienced by a swimming cell during a run. We plot the difference between ramped up and ramped down cases in Fig \ref{fig:ftum}b and find a peak at the optimal cluster size. 

From our data in Figs. \ref{fig:pxv} and \ref{fig:ftum} it follows that for various different performance criteria, the single cell chemotaxis shows optimality. The position of the performance peak may slightly vary depending on the specific measure we use to quantify performance, as seen in the above plots. But the most striking feature here is the existence of a peak at some value of $n$. Below we explain the origin behind this effect.

\begin{figure}
\includegraphics[scale=1.2]{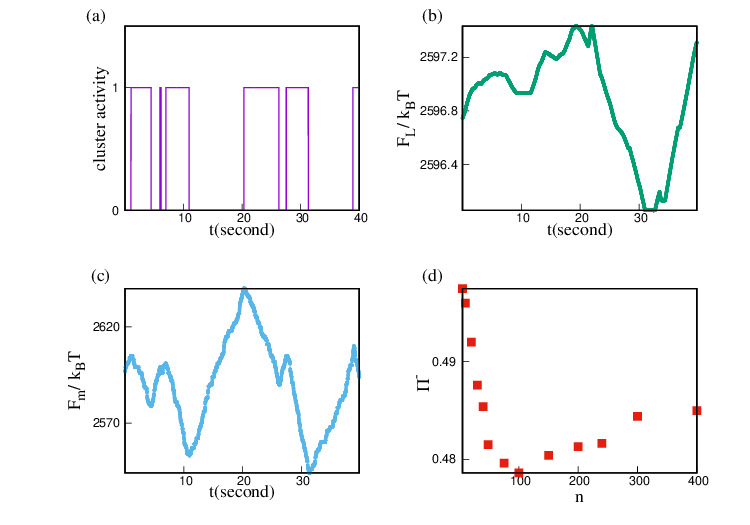}
\caption{Typical time-series of activity along with methylation component and ligand component of free energy of a receptor cluster of size $n=200$. The time-series has been recorded in steady state over a time-window of $40$ $s$. (a) A few transitions of activity state of the cluster. (b) Simultaneous variation of free energy (in unit of $k_{B}T$) due to ligand binding which directly captures the run-tumble trajectory of the cell. (c) Variation of methylation free energy (in unit of $k_{B}T$) of the cluster which is seen to roughly follow the activity transitions.  The scale of variation of ligand binding energy is negligible compared to that of methylation for the present value of $n$. (d) Probability $\Pi^-$ that in a time interval $T=40$ $s$ the net displacement of the cell is negative, shows a minimum and then increases for large $n$. The simulation parameters used are same as those in Fig. \ref{fig:pxv}b main plot.}
\label{fig:aft}
\end{figure}

\begin{figure}
\includegraphics[scale=1.2]{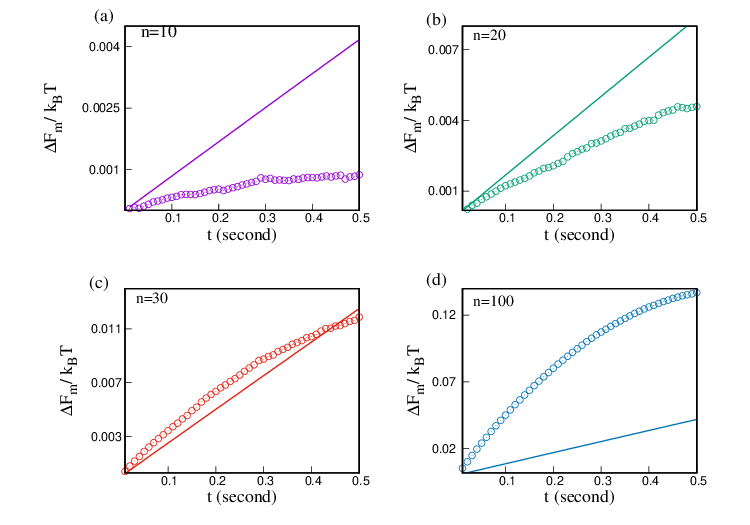}
\caption{Average change $\Delta F_m$ (in unit of $k_{B}T$) in methylation free energy (discrete points) of a cluster for  first $0.5$ $s$ during an uphill run for four different $n$ values. The continuous lines show the change in ligand free energy (in unit of $k_{B}T$) of the cluster. For small $n$ the change in ligand free energy dominates but as $n$ increases, $\Delta F_m$ takes over. These data have been averaged over at least  $2 \times 10^6$ histories.}
\label{fig:delft}
\end{figure}

{\it Competition between sensing and adaptation:} The probability to find a receptor cluster in the active state is $[1+\exp(F_L - F_m)]^{-1}$, where $F_L$ is the sum of ligand binding energy of all dimers in the cluster and $F_m$ is the total methylation of all those dimers. Since the contribution due to ligand binding is the same for all dimers, $F_L$ is proportional to $n$. As the cell swims up (down) the ligand concentration gradient, $F_L$ increases (decreases) with time [see Eq. (\ref{eq:ep})] and this change is proportionately larger with $n$. This means as $n$ increases, the activity of a receptor cluster decreases (increases) quickly during an uphill (downhill) run, thereby elongating (shortening) the run, since activity controls the tumbling rate (see model details in Ref. \cite{sup_mat}). This is why the chemotactic performance gets better with $n$. For large $n$, however, the number of clusters is less and the activity which is calculated by averaging over all clusters, now shows large fluctuations. Switching the activity state of one large cluster brings about large change in the total activity of the cell. For example, when the activity gets too low, all the inactive dimers in a cluster tend to get methylated. This increases $F_m$ significantly and the change in $F_m$ overrides the change in $F_L$. See Fig. \ref{fig:aft} for a typical time-series of cluster activity, $F_m$ and $F_L$ for a large $n$ value. In Fig. \ref{fig:delft} we plot the average change $\Delta F_m$ in methylation free energy  as a function of time during an uphill run of the cell for various different $n$. The change in $F_L$ has been shown with a continuous line for reference. These plots clearly show for large $n$ the change in $F_m$ overtakes the change in $F_L$. The variation in cluster free energy is then controlled by $F_m$. The cell is now less sensitive to the ligand concentration profile. A shorter uphill run and a longer downhill run now become increasingly likely. In Fig. \ref{fig:aft}d we plot the probability to find a negative net displacement of the cell during a time interval $T$ and indeed after reaching a minimum this probability increases again for large $n$. This reduces the value of the chemotactic drift velocity and brings down the performance when $n$ is large. 

Throughout this work, we have considered weak spatial gradient of the nutrient (see Table S1 in Ref. \cite{sup_mat}). Even in the presence of a strong gradient our main conclusions remain valid and we find the optimal cluster size is somewhat larger when the gradient is stronger (data shown in Fig. S5 in Ref. \cite{sup_mat}). Although rapid spatial variation of nutrient has been shown to induce large activity fluctuations in the cell \cite{long2017feedback, sun2017macroscopic, micali2017drift, xue2016moment} which is expected to trigger a stronger methylation response, the ligand free energy changes by a larger amount when the cell is running through a rapidly varying nutrient profile which shifts the trade-off point towards a larger cluster size.

{\it Conclusions:} In this work we have investigated the role of receptor clustering on the chemotactic efficiency of a cell. Although receptor cooperativity amplifies the cell sensitivity towards small variation in nutrient level, the activity fluctuations inside the cell also increase since the number of clusters go down. Large deviation of activity from its mean value triggers large change in methylation levels to ensure adaptation in the biochemical network. The ligand binding energy cannot keep up with such large change in methylation energy and the free energy difference between the active and inactive states gets controlled by methylation now. The above interplay gives rise to a performance peak at an intermediate value of the receptor cluster size.

For a noisy nutrient environment, an optimal size of the receptor signaling team was reported in earlier studies \cite{endres2008variable, aquino2011optimal}. It was argued that receptor cooperativity amplifies not only the ligand signal, but also the noise present in it. For optimal performance, therefore, a trade-off is required where the signaling team size should be large enough for sensitive detection of small changes in ligand concentration, but small enough such that the amplified noise does not insensitize the cell response \cite{endres2008variable}. Moreover, when both ligand noise and intracellular biochemical noise are considered, the receptor clustering is beneficial as long as the amplified ligand noise stays below the biochemical noise \cite{aquino2011optimal}. On the other hand, we find an optimal team size even when the ligand concentration profile does not fluctuate with time, and the origin of this optimality to our best knowledge has not been reported previously.

It should be possible to test our results qualitatively in experiment. Both for a swimming cell and tethered cell we have observed the best chemotactic performance at a specific size of the receptor cluster. In our model the best performance is observed for clusters which contain $\sim 70$ TDs. However, it may not be possible to find accurate quantitative agreement between our model and experiments. To keep our model simple and tractable, we have not considered a few aspects of the intracellular reaction network, like hexagonal geometry of the spatial arrangement of the receptor array \cite{briegel2012bacterial, liu2012molecular}, or more importantly, the energy cost due to curvature of the cell membrane induced by the receptor clusters \cite{endres2009polar, haselwandter2014role, draper2017origins}. But our main conclusions should not get affected by these assumptions and the interplay between ligand free energy and methylation free energy can be experimentally investigated as the cooperative interaction among the receptors is varied \cite{colin2017multiple, keegstra2017phenotypic}. A stronger interaction among the receptors which is responsible for formation of larger clusters, has been experimentally shown to induce larger activity fluctuations in a tethered cell \cite{colin2017multiple, keegstra2017phenotypic}. Whether the variation of methylation free energy increases for stronger receptor interaction and its effect on the chemotactic efficiency, $(\tau_\uparrow - \tau_\downarrow )$ (Fig. \ref{fig:ftum}b) can be investigated in experiments.  Finally, our study opens up the important question of competition between sensing and adaptation, which is relevant in a wide variety of biological systems \cite{gepshtein2013sensory, solomon2014moving, wark2007sensory}. It would be interesting to see if this competition gives rise to similar performance peaks in this broad class of systems as well.

\section*{Acknowledgements} 
SDM acknowledges a research fellowship [Grant No. 09/575(0122)/2019-EMR -I]  from the Council of Scientific and Industrial Research (CSIR), India. SC acknowledges financial support from the Science and Engineering Research Board, India (Grant No: MTR/2019/000946).


\begin{thebibliography}{10}
\bibitem{sourjik2002receptor}
Victor Sourjik and Howard~C Berg.
\newblock Receptor sensitivity in bacterial chemotaxis.
\newblock {\em Proceedings of the National Academy of Sciences},
  99(1):123--127, 2002.

\bibitem{turner2000real}
Linda Turner, William~S Ryu, and Howard~C Berg.
\newblock Real-time imaging of fluorescent flagellar filaments.
\newblock {\em Journal of bacteriology}, 182(10):2793--2801, 2000.

\bibitem{elowitz2002stochastic}
Michael~B Elowitz, Arnold~J Levine, Eric~D Siggia, and Peter~S Swain.
\newblock Stochastic gene expression in a single cell.
\newblock {\em Science}, 297(5584):1183--1186, 2002.

\bibitem{raj2008nature}
Arjun Raj and Alexander Van~Oudenaarden.
\newblock Nature, nurture, or chance: stochastic gene expression and its
  consequences.
\newblock {\em Cell}, 135(2):216--226, 2008.

\bibitem{rao2002control}
Christopher~V Rao, Denise~M Wolf, and Adam~P Arkin.
\newblock Control, exploitation and tolerance of intracellular noise.
\newblock {\em Nature}, 420(6912):231--237, 2002.

\bibitem{lan2016information}
Ganhui Lan and Yuhai Tu.
\newblock Information processing in bacteria: memory, computation, and
  statistical physics: a key issues review.
\newblock {\em Reports on Progress in Physics}, 79(5):052601, 2016.

\bibitem{berg2008coli}
Howard~C Berg.
\newblock {\em E. coli in Motion}.
\newblock Springer Science \& Business Media, Heidelberg, 2008.

\bibitem{frank2016networked}
Vered Frank, Germ{\'a}n~E Pi{\~n}as, Harel Cohen, John~S Parkinson, and Ady
  Vaknin.
\newblock Networked chemoreceptors benefit bacterial chemotaxis performance.
\newblock {\em MBio}, 7(6):e01824, 2016.

\bibitem{duke1999heightened}
TAJ Duke and Dennis Bray.
\newblock Heightened sensitivity of a lattice of membrane receptors.
\newblock {\em Proceedings of the National Academy of Sciences},
  96(18):10104--10108, 1999.

\bibitem{bray1998receptor}
Dennis Bray, Matthew~D Levin, and Carl~J Morton-Firth.
\newblock Receptor clustering as a cellular mechanism to control sensitivity.
\newblock {\em Nature}, 393(6680):85--88, 1998.

\bibitem{colin2017multiple}
Remy Colin, Christelle Rosazza, Ady Vaknin, and Victor Sourjik.
\newblock Multiple sources of slow activity fluctuations in a bacterial
  chemosensory network.
\newblock {\em Elife}, 6:e26796, 2017.

\bibitem{keegstra2017phenotypic}
Johannes~M Keegstra, Keita Kamino, Fran{\c{c}}ois Anquez, Milena~D Lazova,
  Thierry Emonet, and Thomas~S Shimizu.
\newblock Phenotypic diversity and temporal variability in a bacterial
  signaling network revealed by single-cell fret.
\newblock {\em Elife}, 6:e27455, 2017.

\bibitem{korobkova2004molecular}
Ekaterina Korobkova, Thierry Emonet, Jose~MG Vilar, Thomas~S Shimizu, and
  Philippe Cluzel.
\newblock From molecular noise to behavioural variability in a single
  bacterium.
\newblock {\em Nature}, 428(6982):574--578, 2004.

\bibitem{matthaus2011origin}
Franziska Matth{\"a}us, Mario~S Mommer, Tine Curk, and Jure Dobnikar.
\newblock On the origin and characteristics of noise-induced l{\'e}vy walks of
  e. coli.
\newblock {\em PloS one}, 6(4):e18623, 2011.

\bibitem{flores2012signaling}
Marlo Flores, Thomas~S Shimizu, Pieter~Rein ten Wolde, and Filipe Tostevin.
\newblock Signaling noise enhances chemotactic drift of e. coli.
\newblock {\em Physical review letters}, 109(14):148101, 2012.

\bibitem{dev2018optimal}
Subrata Dev and Sakuntala Chatterjee.
\newblock Optimal methylation noise for best chemotactic performance of e.
  coli.
\newblock {\em Physical Review E}, 97(3):032420, 2018.

\bibitem{tu2005white}
Yuhai Tu and G~Grinstein.
\newblock How white noise generates power-law switching in bacterial flagellar
  motors.
\newblock {\em Physical review letters}, 94(20):208101, 2005.

\bibitem{sup_mat}
See supplementary materials here for details of the model, simulation
  technique, list of paramter values and additional data.

\bibitem{dev2015optimal}
Subrata Dev and Sakuntala Chatterjee.
\newblock Optimal search in e. coli chemotaxis.
\newblock {\em Physical Review E}, 91(4):042714, 2015.

\bibitem{liu2012molecular}
Jun Liu, Bo~Hu, Dustin~R Morado, Sneha Jani, Michael~D Manson, and William
  Margolin.
\newblock Molecular architecture of chemoreceptor arrays revealed by
  cryoelectron tomography of escherichia coli minicells.
\newblock {\em Proceedings of National Academy of Sciences},
  109(23):E1481--E1488, 2012.

\bibitem{briegel2012bacterial}
Ariane Briegel, Xiaoxiao Li, Alexandrine~M Bilwes, Kelly~T Hughes, Grant~J
  Jensen, and Brian~R Crane.
\newblock Bacterial chemoreceptor arrays are hexagonally packed trimers of
  receptor dimers networked by rings of kinase and coupling proteins.
\newblock {\em Proceedings of the National Academy of Sciences},
  109(10):3766--3771, 2012.

\bibitem{mello2005allosteric}
Bernardo~A Mello and Yuhai Tu.
\newblock An allosteric model for heterogeneous receptor complexes:
  understanding bacterial chemotaxis responses to multiple stimuli.
\newblock {\em Proceedings of the National Academy of Sciences},
  102(48):17354--17359, 2005.

\bibitem{keymer2006chemosensing}
Juan~E Keymer, Robert~G Endres, Monica Skoge, Yigal Meir, and Ned~S Wingreen.
\newblock Chemosensing in escherichia coli: two regimes of two-state receptors.
\newblock {\em Proceedings of the National Academy of Sciences},
  103(6):1786--1791, 2006.

\bibitem{monod1965nature}
Jacque Monod, Jeffries Wyman, and Jean-Pierre Changeux.
\newblock On the nature of allosteric transitions: a plausible model.
\newblock {\em J Mol Biol}, 12(1):88--118, 1965.

\bibitem{li2004cellular}
Mingshan Li and Gerald~L Hazelbauer.
\newblock Cellular stoichiometry of the components of the chemotaxis signaling
  complex.
\newblock {\em Journal of bacteriology}, 186(12):3687--3694, 2004.

\bibitem{schulmeister2008protein}
Sonja Schulmeister, Michaela Ruttorf, Sebastian Thiem, David Kentner, Dirk
  Lebiedz, and Victor Sourjik.
\newblock Protein exchange dynamics at chemoreceptor clusters in escherichia
  coli.
\newblock {\em Proceedings of the National Academy of Sciences},
  105(17):6403--6408, 2008.

\bibitem{berg1975transient}
Howard~C Berg and PM~Tedesco.
\newblock Transient response to chemotactic stimuli in escherichia coli.
\newblock {\em Proceedings of the National Academy of Sciences},
  72(8):3235--3239, 1975.

\bibitem{goy1977sensory}
Michael~F Goy, Martin~S Springer, and Julius Adler.
\newblock Sensory transduction in escherichia coli: role of a protein
  methylation reaction in sensory adaptation.
\newblock {\em Proceedings of the National Academy of Sciences},
  74(11):4964--4968, 1977.

\bibitem{levin2002binding}
Matthew~D Levin, Thomas~S Shimizu, and Dennis Bray.
\newblock Binding and diffusion of cher molecules within a cluster of membrane
  receptors.
\newblock {\em Biophysical journal}, 82(4):1809--1817, 2002.

\bibitem{endres2006precise}
Robert~G Endres and Ned~S Wingreen.
\newblock Precise adaptation in bacterial chemotaxis through “assistance
  neighborhoods”.
\newblock {\em Proceedings of the National Academy of Sciences},
  103(35):13040--13044, 2006.

\bibitem{hansen2008chemotaxis}
Clinton~H Hansen, Robert~G Endres, and Ned~S Wingreen.
\newblock Chemotaxis in escherichia coli: a molecular model for robust precise
  adaptation.
\newblock {\em PLoS Comput Biol}, 4(1):e1, 2008.

\bibitem{kim2002dynamic}
Sung-Hou Kim, Weiru Wang, and Kyeong~Kyu Kim.
\newblock Dynamic and clustering model of bacterial chemotaxis receptors:
  structural basis for signaling and high sensitivity.
\newblock {\em Proceedings of the National Academy of Sciences},
  99(18):11611--11615, 2002.

\bibitem{li2005adaptational}
Mingshan Li and Gerald~L Hazelbauer.
\newblock Adaptational assistance in clusters of bacterial chemoreceptors.
\newblock {\em Molecular microbiology}, 56(6):1617--1626, 2005.

\bibitem{le1997methylation}
Herv{\'e} Le~Moual, Tony Quang, and Daniel~E Koshland.
\newblock Methylation of the escherichia coli chemotaxis receptors: intra-and
  interdimer mechanisms.
\newblock {\em Biochemistry}, 36(43):13441--13448, 1997.

\bibitem{celani2010bacterial}
Antonio Celani and Massimo Vergassola.
\newblock Bacterial strategies for chemotaxis response.
\newblock {\em Proceedings of the National Academy of Sciences},
  107(4):1391--1396, 2010.

\bibitem{de2004chemotaxis}
P-G De~Gennes.
\newblock Chemotaxis: the role of internal delays.
\newblock {\em European Biophysics Journal}, 33(8):691--693, 2004.

\bibitem{chatterjee2011chemotaxis}
Sakuntala Chatterjee, Rava~Azeredo da~Silveira, and Yariv Kafri.
\newblock Chemotaxis when bacteria remember: drift versus diffusion.
\newblock {\em PLoS computational biology}, 7(12), 2011.

\bibitem{pankratova2018chemotactic}
Evgeniya~V Pankratova, Alena~I Kalyakulina, Mikhail~I Krivonosov, Sergei~V
  Denisov, Katja~M Taute, and Vasily~Yu Zaburdaev.
\newblock Chemotactic drift speed for bacterial motility pattern with two
  alternating turning events.
\newblock {\em PloS one}, 13(1):e0190434, 2018.

\bibitem{samanta2017predicting}
Sibendu Samanta, Ritwik Layek, Shantimoy Kar, M~Kiran Raj, Sudipta
  Mukhopadhyay, and Suman Chakraborty.
\newblock Predicting escherichia coli's chemotactic drift under exponential
  gradient.
\newblock {\em Physical Review E}, 96(3):032409, 2017.

\bibitem{long2017feedback}
Junjiajia Long, Steven~W Zucker, and Thierry Emonet.
\newblock Feedback between motion and sensation provides nonlinear boost in
  run-and-tumble navigation.
\newblock {\em PLoS computational biology}, 13(3):e1005429, 2017.

\bibitem{sun2017macroscopic}
Weiran Sun and Min Tang.
\newblock Macroscopic limits of pathway-based kinetic models for e. coli
  chemotaxis in large gradient environments.
\newblock {\em Multiscale Modeling \& Simulation}, 15(2):797--826, 2017. DOI: 10.1137/16M1074011

\bibitem{micali2017drift}
Gabriele Micali, R{\'e}my Colin, Victor Sourjik, and Robert~G Endres.
\newblock Drift and behavior of e. coli cells.
\newblock {\em Biophysical journal}, 113(11):2321--2325, 2017.

\bibitem{xue2016moment}
Chuan Xue and Xige Yang.
\newblock Moment-flux models for bacterial chemotaxis in large signal
  gradients.
\newblock {\em Journal of mathematical biology}, 73(4):977--1000, 2016.

\bibitem{endres2008variable}
Robert~G Endres, Olga Oleksiuk, Clinton~H Hansen, Yigal Meir, Victor Sourjik,
  and Ned~S Wingreen.
\newblock Variable sizes of escherichia coli chemoreceptor signaling teams.
\newblock {\em Molecular systems biology}, 4(1):211, 2008.

\bibitem{aquino2011optimal}
Gerardo Aquino, Diana Clausznitzer, Sylvain Tollis, and Robert~G Endres.
\newblock Optimal receptor-cluster size determined by intrinsic and extrinsic
  noise.
\newblock {\em Physical Review E}, 83(2):021914, 2011.

\bibitem{endres2009polar}
Robert~G Endres.
\newblock Polar chemoreceptor clustering by coupled trimers of dimers.
\newblock {\em Biophysical journal}, 96(2):453--463, 2009.

\bibitem{haselwandter2014role}
Christoph~A Haselwandter and Ned~S Wingreen.
\newblock The role of membrane-mediated interactions in the assembly and
  architecture of chemoreceptor lattices.
\newblock {\em PLoS computational biology}, 10(12):e1003932, 2014.

\bibitem{draper2017origins}
Will Draper and Jan Liphardt.
\newblock Origins of chemoreceptor curvature sorting in escherichia coli.
\newblock {\em Nature communications}, 8(1):1--9, 2017.

\bibitem{gepshtein2013sensory}
Sergei Gepshtein, Luis~A Lesmes, and Thomas~D Albright.
\newblock Sensory adaptation as optimal resource allocation.
\newblock {\em Proceedings of the National Academy of Sciences},
  110(11):4368--4373, 2013.

\bibitem{solomon2014moving}
Samuel~G Solomon and Adam Kohn.
\newblock Moving sensory adaptation beyond suppressive effects in single
  neurons.
\newblock {\em Current Biology}, 24(20):R1012--R1022, 2014.

\bibitem{wark2007sensory}
Barry Wark, Brian~Nils Lundstrom, and Adrienne Fairhall.
\newblock Sensory adaptation.
\newblock {\em Current opinion in neurobiology}, 17(4):423--429, 2007.
\end{thebibliography}
		
\end{document}




\newcommand{\be}{\begin{equation}}
\newcommand{\ee}{\end{equation}}
\newcommand{\bea}{\begin{eqnarray}}
\newcommand{\eea}{\end{eqnarray}}
\newcommand{\bi}{\begin{itemize}}
\newcommand{\ei}{\end{itemize}}

\section{Complete model description and simulation details}

There are three major steps in our simulation of the chemotactic reaction network and cell motion. These are 
\bi 
\item Activity switching of the receptor clusters
\item Binding/unbinding dynamics of CheR and CheB-P enzymes to the receptor dimers and modification of dimer methylation levels by these enzymes
\item Calculation of CheY-P level from total activity and the resulting run-and-tumble motion of the cell. 
\ei

Below we describe each of these steps in details.

A receptor cluster containing $n$ trimers of dimers can switch between active state ($a=1$) and inactive state ($a=0$). The free energy difference between the two states is given by the Monod-Wyman-Changeux model \cite{monod1965nature}
\be
F = 3n\left ( 1+ \log \frac{1+c(x)/K_{min}}{1+c(x)/K_{max}} \right ) - \sum_{i=1}^{3n} m_i
\ee
where $K_{min}$ and $K_{max}$ set the range of sensitivity of the cell, {\sl i.e.} the cell can sense a ligand concentration level $c(x)$ for $K_{min} < c(x) < K_{max}$. We have used $K_{max} = 3 mM$ and two values of $K_{min}$ which are $7 \mu M$ \cite{colin2017multiple} and $18 \mu M$ \cite{flores2012signaling}. We do not find any significant difference between these two choices. In our model we do not explicitly consider the binding events of the nutrient molecules to the receptors and assume all dimers experience a mean nutrient level $c(x)$ that depends on the current cell position. The methylation level of the $i$-th dimer in the cluster is $m_i$ which can take any integer value between $0$ and $8$. A completely demethylated dimer has $m_i=0$ and a completely methylated dimer has $m_i=8$.

The probability to find a receptor cluster in active state is $[1+\exp(F)]^{-1}$. The transition rate between the two activity states is chosen based on local detailed balance \cite{colin2017multiple}. From $a=0$ state the receptor cluster switches to $a=1$ state with rate $\dfrac{w_a}{1+\exp(F)}$ and the reverse switch happens with rate $\dfrac{w_a \exp(F)}{1+\exp(F)}$. Here, $w_a$ is the characteristic time-scale of the transition. Here we have presented data for $w_a = 0.75/s$ which is a bit higher than the value $w_a=0.25/s$ used in \cite{colin2017multiple} to explain the experimental results. We have verified (data not shown here) that no significant change results from this difference.

We consider a total of $140$ CheR molecules and $240$ CheB molecules in the cell \cite{li2004cellular}. An unbound CheR molecule which is in the cell cytoplasm can bind to a receptor dimer, provided no other enzyme is bound to it. The binding can take place at the receptor tether or modification site \cite{pontius2013adaptation, wu1996receptor, feng1999enhanced}. Although both these types of binding are slow, the binding at the tether is relatively faster than that at the modification site \cite{pontius2013adaptation, schulmeister2008protein}. Because of this we assume in our model that an unbound CheR only binds to the tether of the receptor with binding rate $w_r$. Once bound, the CheR enzyme can raise the methylation level of the dimer  with rate $k_{r}$ provided the dimer belongs to an inactive cluster and its methylation level is less than $8$. With rate $w_u$ CheR can unbind from the dimer and can either reattach to another unoccupied dimer within the same cluster, or return to the cytoplasmic bulk.

Note that due to very slow binding of the enzyme to the receptors, if one binding results in only one methylation event, then it becomes difficult for the system to maintain perfect adaptation \cite{pontius2013adaptation}. To circumvent this, many different mechanisms were proposed, including assistance neighborhood \cite{hansen2008chemotaxis, li2005adaptational, endres2006precise}  and brachiation \cite{levin2002binding}. In assistance neighborhood model, a bound enzyme can modify the methylation level of neighboring receptors and in brachiation mechanism a bound enzyme can perform random walk on the receptor array before unbinding from it. Both these methods involve methylation of multiple receptors from a single binding event. We incorporate this aspect in our model by allowing rebinding of CheR to other dimers within the same cluster. This step is simple and effective in our model, since we do not explicitly include spatial geometry of the receptor cluster.

A CheB molecule in the cytoplasmic bulk can undergo phosphorylation by an active receptor with the rate $w_p$. In the phosphorylated state, an unbound CheB-P molecule can undergo dephosphorylation with the rate $w_{dp}$. The binding-unbinding dynamics of a CheB-P molecule is very similar to what has been described above for CheR, with the tether binding rate $w_b$. In the bound state a CheB-P molecule can demethylate an active receptor dimer with rate $k_{b}$ provided its methylation level is non-zero. We implement the rebinding process in this case too in the same way as mentioned above.

The CheY-P level depends on the total activity of all the receptors, which is calculated as $A=\sum_k a_k / N_c $ where $a_k$ is the activity of the $k-th$ cluster and $N_c$ is the total number of clusters in the cell. In our simple model we assume all clusters are of equal size, and hence $N_c = N_{dim}/3n$. Therefore, $A$ measures the fraction of active receptor clusters in the cell. Define $Y_P = \dfrac{[\text{CheY-P}]}{[\text{CheY}]}$ and the rate equation that governs $Y_P$ is \cite{flores2012signaling} 
\be 
\frac{dY_P}{dt} = K_Y A (1-Y_P) - K_Z Y_P
\ee
where $K_Y$ and $K_Z$ are rate constants for phosphorylation and dephosphorylation, respectively. In our simulation we directly use $Y_P = \dfrac{A}{A + K_Z/K_Y}$. The run-tumble motility of the cell is controlled by $Y_P$. If the cell is in the run mode, it can switch to tumble mode with rate $\omega \exp (-G)$ where $G = \Delta_1 - \dfrac{\Delta_2}{1+Y_0/Y_P}$ and the opposite switch from tumble to run happens with the rate $\omega \exp(G)$.

In our simulations we consider motion of the cell in one and two spatial dimensions. In the one dimensional case, the cell can either move to the left or to the right with speed $v$ during a run and stays put in the same spot during a tumble. The smallest time-step in our simulation is $dt=0.01s$ during which a running cell can move by an amount $vdt$ which which sets the lattice spacing in our one dimensional model. We set up a nutrient concentration profile $c(x) = c_0 (1+x/x_0)$ in a region of length $L$ and when the cell reaches the boundaries of this region at $x=0,L$, it gets reflected back. For a two dimensional motion of the cell we consider an $L_x \times L_y$ box with reflecting boundaries at the four walls. The nutrient concentration gradient is present along $x$-direction, as in one dimension, and along $y$ direction the concentration is homogeneous. Due to rotational diffusion the trajectory of the cell does not remain a perfect straight line during a run. It shows gradual bending. After each tumble the cell chooses a random direction to start a new run. We measure all response functions in the long time limit when the system has reached a steady state.

\newpage
\section{Model parameters}

\begin{center}
\begin{table}[h]
\caption{}
\begin{tabular}{|l|l|l|l|}
\hline
\textbf{Symbol} & \hspace{30mm} \textbf{Description} & \textbf{Value} & \textbf{References} \\
\hline
$N_{dim}$ & Total number of receptor dimers & $7200$ & \cite{pontius2013adaptation,li2004cellular} \\
\hline
$N_R$  & Total number of CheR  protein molecules &  $140$ & \cite{pontius2013adaptation,li2004cellular}\\
\hline
$N_B$  & Total number of CheB  protein molecules &  $240$ & \cite{pontius2013adaptation,li2004cellular} \\
\hline
$K_{min}$ & Minimum concentration receptor can sense &  $7$ $\mu M$, $18$ $\mu M$ & \cite{colin2017multiple}, \cite{flores2012signaling} \\
\hline
$K_{max}$ & Maximum concentration receptor can sense &  $3000$ $\mu M$ & \cite{flores2012signaling,jiang2010quantitative} \\
\hline
$w_{a}$  & Flipping rate of activity &  $0.75$ $s^{-1}$ & Present study\\
\hline
$\omega$  & Switching frequency of motor &  $1.3$ $s^{-1}$ & \cite{sneddon2012stochastic} \\
\hline
 $\Delta_{1}$  & Nondimensional constant regulating motor switching  &  $10$ & \cite{sneddon2012stochastic} \\
\hline
$\Delta_{2}$  & Nondimensional constant regulating motor switching &  $20$ & \cite{sneddon2012stochastic}  \\
\hline
$Y_{0}$ & Adopted value of the fraction of CheY-P protein &  $0.34$ & \cite{sneddon2012stochastic} \\
\hline 
$K_{Y}$ & Phosphorylation rate of CheY molecule &  $1.7$ $s^{-1}$ & \cite{flores2012signaling,tu2008modeling} \\
\hline
$K_{Z}$ & Dephosphorylation rate of CheY molecule &  $2$ $s^{-1}$ & \cite{flores2012signaling,tu2008modeling}\\
\hline
$w_{r}$ & Binding rate of bulk CheR to tether site of an unoccupied dimer &  $0.068$ $s^{-1}$ & \cite{pontius2013adaptation,schulmeister2008protein} \\
\hline
$w_{b}$ & Binding rate of bulk CheB-P to tether site of an unoccupied dimer &  $0.061$ $s^{-1}$ & \cite{pontius2013adaptation,schulmeister2008protein} \\
\hline
$w_{u}$ & Unbinding rate of bound CheR and CheB-P &  $5$ $s^{-1}$ & \cite{pontius2013adaptation,schulmeister2008protein} \\	
\hline	
$k_{r}$ & Methylation rate of bound CheR & $2.7$ $s^{-1}$ & \cite{pontius2013adaptation,schulmeister2008protein}\\ 	
\hline
$k_{b}$ & Demethylation rate of bound CheB-P &  $3$ $s^{-1}$ & \cite{pontius2013adaptation,schulmeister2008protein}\\
\hline	
$w_{p}$ & CheB phosphorylation rate &  $3$ $s^{-1}$ & \cite{pontius2013adaptation,stewart2000rapid} \\
\hline
$w_{dp}$ & CheB-P dephosphorylation rate &  $0.37$ $s^{-1}$ & \cite{pontius2013adaptation}\\	
\hline	
$L$ &  Box length in 1D &  $2000$ $\mu m$ & Present study \\
\hline
$v$ & Speed of the cell &  $20$ $\mu m/s$ & \cite{berg2008coli} \\
\hline
$dt$ & Time step &  $0.01$ $s$ & Present study\\
\hline
$L_{x}\times L_{y}$ & Box dimension in 2D&  $2000\times 200$ $\mu m^{2}$ & Present study \\
\hline
$D_{\Theta}$ & Rotational Diffusivity &  $0.062$ $\mu m^{2}/s$ & \cite{berg1972chemotaxis,dufour2014limits,karmakar2016enhancement}\\
\hline
$c_{0}$ & Background nutrient concentration &  $200$ $\mu M$ & Present study \\
\hline
$1/x_{0}$ & Linear concentration gradient of nutrient&  $0.00005$ $\mu m^{-1}$ & Present study \\
\hline
\end{tabular}
\label{table}
\end{table}
\end{center} 	

\newpage
\section{Activity fluctuations increase with receptor clustering}

\begin{figure}[h!]      
\includegraphics[scale=1.0]{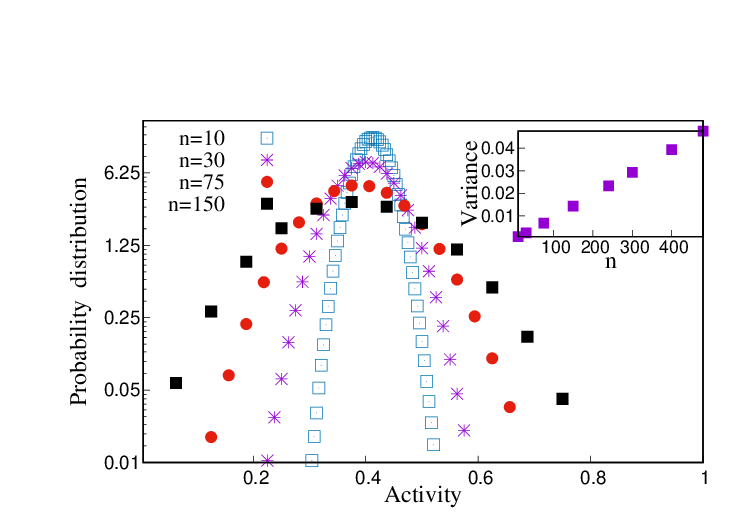}
\caption{Probability density of activity (defined as the fraction of active clusters) for few representative values of receptor cluster size $n$. As $n$ increases, distribution gets wider, which is consistent with the experimental observation in \cite{colin2017multiple, keegstra2017phenotypic}.  Inset shows variance of the distribution scales linearly with $n$.}   
\label{fig:ndm3}
\end{figure}

 The total number of receptor dimers $N_{dim}$ is fixed in a cell. The number of trimers of dimers (TD) then becomes $N_{dim}/3$. In each cluster there are $n$ such TDs and therefore the number of clusters is $N_c=N_{dim}/3n$. As the clusters become bigger, $N_c$ decreases. Each cluster can be either in an active state or in an inactive state. Let $a_k(t)$ denote a stochastic variable which takes the value $1$ if the $k$-th cluster is active at time $t$ or $0$ if it is inactive. The total activity of the cell is then defined as $\sum_{k=1}^{N_c} a_k(t) / N_c$, which gives the fraction of active clusters at time $t$. If we assume the activity switching for different clusters occur independently, then from central limit theorem it follows that total activity will have a Gaussian probability distribution whose width scales as $n$. In our model, $a_k(t)$ for different $k$ may not be completely uncorrelated, but we do find from our simulations that fluctuations in activity increases linearly with $n$ as shown in the inset of \ref{fig:ndm3}. 

\newpage
\section{Results in one dimension}

\begin{figure}[h!]
\includegraphics[scale=1.6]{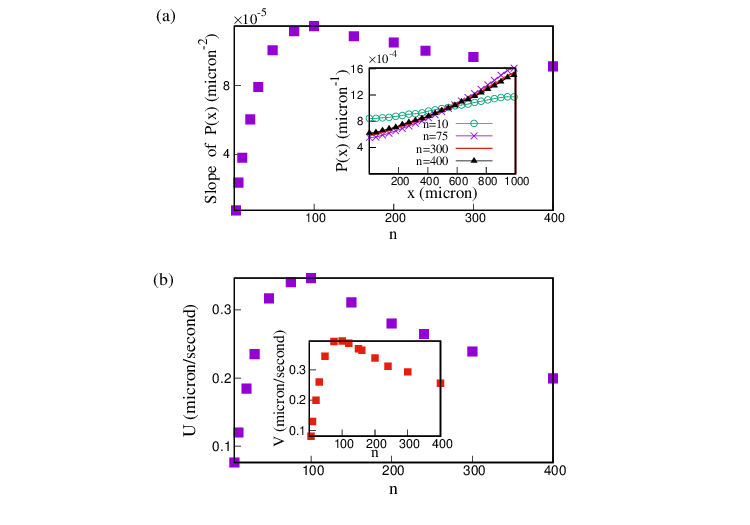}
\caption{(a) In presence of a linear nutrient concentration profile with weak gradient, the steady state position distribution $P(x)$ of the cell is almost linear. The steepness shows a peak with $n$. Inset shows some representative $P(x)$ plots for few chosen $n$ values. (b) Chemotactic drift velocity measured from net displacement in a run (main plot) and from net displacement in a fixed time-interval $T=10$ $s$ (inset) show peak with $n$. The data points are averaged over at least $10^7$ histories. The simulation parameters are listed in Table \ref{table}. }   
\label{fig:pxv1}
\end{figure}

\begin{figure}[h!]         
\includegraphics[scale=1]{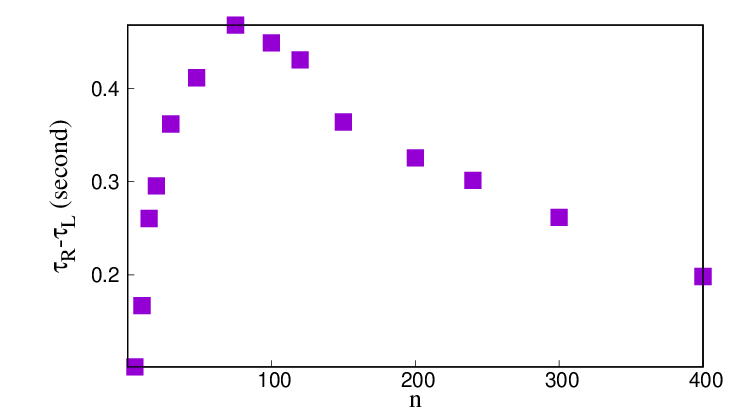}
\caption{During a rightward (leftward) run of the cell when it is headed towards regions with more (less) nutrient, $\tau_R$ ($\tau_L$) denote the average time till the next tumble. The difference  $(\tau_{R}-\tau_{L})$ shows a peak as a function of $n$. Each data point is averaged over at least $10^6$ histories. All simulation parameters are as in Fig. \ref{fig:pxv1}}.
\label{fig:ndm1}
\end{figure}

\begin{figure}[h!]         
\includegraphics[scale=1.3]{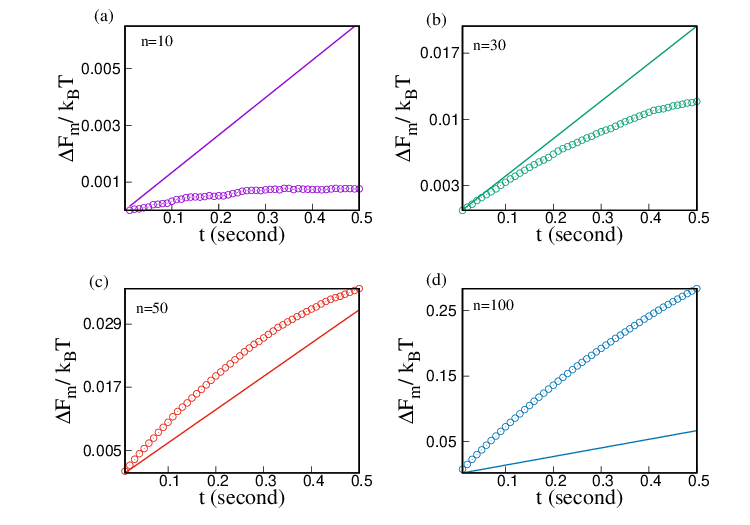}
\caption{Average change $\Delta F_m$ (in unit of $k_{B}T$) in methylation free energy (discrete points) of a cluster for first $0.5$ $s$ during an uphill run for four different cluster sizes. For comparison, the corresponding change in ligand free energy (in unit of $k_{B}T$) has been shown by continuous lines. For small $n$ the change in ligand free energy dominates but as $n$ increases, $\Delta F_m$ takes over. These data have been averaged over at least  $2 \times 10^6$ histories.}   
\label{fig:ndm2}
\end{figure}

In this section, we present our results for the cell motion in one spatial dimension, in presence of a linear $c(x)=c_0 (1
+x/x_0)$, as in the main paper. All our conclusions presented in the main paper for the two dimensional case remain valid here.  Our simulation data are shown in Figs. \ref{fig:pxv1}, \ref{fig:ndm1} and \ref{fig:ndm2}. 

\newpage 
\section{Data for stronger gradient and exponential gradient}
\begin{figure}[h!]
\includegraphics[scale=1.5]{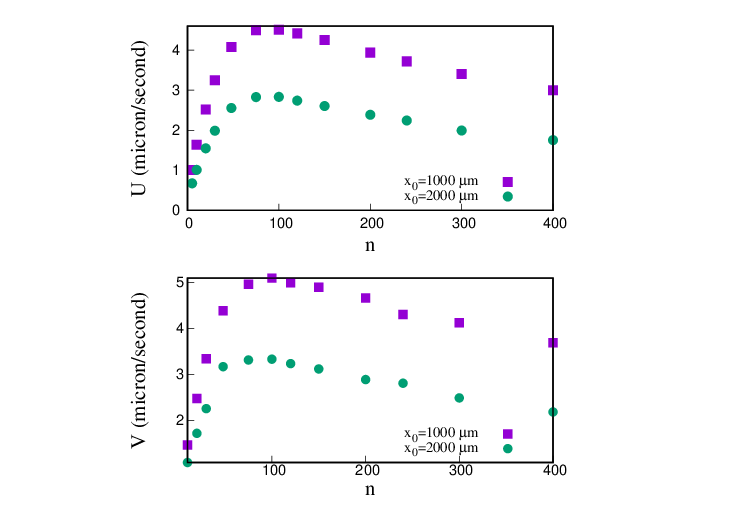}
\caption{ Stronger gradient shifts the performance peak slightly rightward. Here we have used a linear $c(x) = c_0 (1+x/x_0)$ with $x_0=1000$ $\mu m$ (purple square) and $2000$ $\mu m$ (green circle), which are much more steeper than our usual choice of $x_0 =20000$ $\mu m$ (see Table \ref{table}). Other simulation parameters are listed in Table \ref{table}. Each data point has been averaged over at least $10^7$ histories.}   
\label{fig:pxv2}
\end{figure}
\newpage
\begin{figure}[h!]
\includegraphics[scale=1]{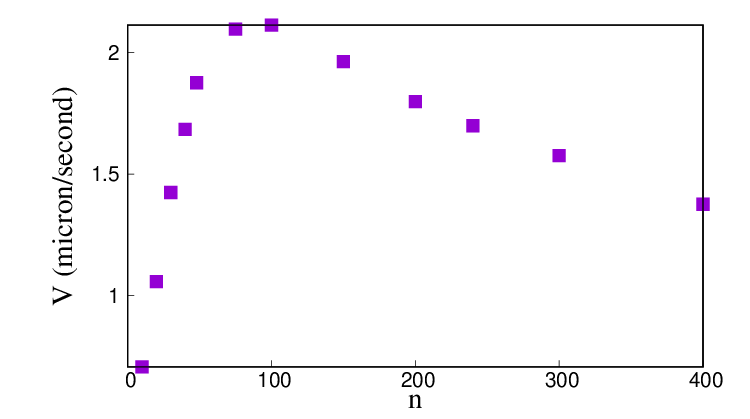}
\caption{ For an exponential nutrient profile, $c(x)=c_0 \exp (x/\xi)$ with $\xi=4000$ $\mu m$, we find a performance peak. The simulation parameters are listed in Table \ref{table}. Each data point has been averaged over at least $10^7$ histories.}    
\label{fig:pxv3}
\end{figure}
